\documentclass[amsmath,preprint,endfloats,prb,12pt]{revtex4}

\usepackage{bm}
\usepackage{amsmath}
\usepackage{graphicx}

\begin{document}

\title{
       Efficient energy transfer in light-harvesting systems, {III}: 
       The influence of the eighth bacteriochlorophyll on 
       the dynamics and efficiency in FMO 
       }

\author{
       Jeremy Moix
       } 
\affiliation{
       Department of Chemistry,
       Massachusetts Institute of Technology,
       77 Massachusetts Avenue, Cambridge, MA 02139
       }
\affiliation{
       Singapore MIT Alliance for Research and Technology,
       18 Medical Drive, Singapore 117456
       }
\affiliation{
       School of Materials Science and Engineering, 
       Nanyang Technological University, 
       Singapore 639798
       }

\author{
       Jianlan Wu
       } 
\affiliation{
       Department of Physics,
       Zhejiang University,
       38 ZheDa Road, Hangzhou, China, 310027
       }
\affiliation{
       Singapore MIT Alliance for Research and Technology,
       18 Medical Drive, Singapore 117456
       }

\author{
       Pengfei Huo
       }
\author{
       David Coker
       }
\affiliation{
       Department of Chemistry, 
       Boston University, 
       590 Commonwealth Avenue, Boston, Massachusetts 02215, USA
       }
\affiliation{
       Department of Physics, 
       University College Dublin, 
       Dublin 4, Ireland 
       }

\author{
       Jianshu Cao
       } \email{jianshu@mit.edu}
\affiliation{
       Department of Chemistry,
       Massachusetts Institute of Technology,
       77 Massachusetts Avenue, Cambridge, MA 02139
       }
\affiliation{
       Singapore MIT Alliance for Research and Technology,
       18 Medical Drive, Singapore 117456
       }

\date{\today}

\begin{abstract}

The most recent crystal structure of the Fenna-Matthews-Olson (FMO) 
protein complex indicates that each subunit contains an 
additional eighth chromophore.
It has been proposed that this extra site functions as a 
link between the chlorosome antenna complex and the remaining seven
chromophores in FMO 
[Schmidt am Busch et al, J. Phys. Chem. Lett., {\bf 2}, 93 (2011)].
Here, we investigate the implications of this scenario 
through numerical calculations with the generalized Bloch-Redfield (GBR) 
equation and the non-interacting blip approximation (NIBA). 
Three key insights into the population dynamics and energy transfer
efficiency in FMO are provided.
First, it is shown that the oscillations that are often observed
in the population relaxation of the dimer composed of sites one and two 
may be completely suppressed in the eight site model.
The presence of the coherent oscillations is shown to depend 
upon the particular initial preparation of the dimer state.
Secondly it is demonstrated that while the presence of the eighth 
chromophore does not cause a dramatic change in the energy transfer efficiency, 
it does however lead to a dominant energy transfer pathway
which can be characterized by an effective three site system arranged 
in an equally-spaced downhill configuration.
Such a configuration leads to an optimal value of the site energy of 
the eighth chromophore which is shown to be near to its suggested value.
Finally we confirm that the energy transfer process in the eight site 
FMO complex remains efficient and robust.
The optimal values of the bath parameters are computed and
shown to be closer to the experimentally fitted values 
than those calculated previously for the seven site system.

\end{abstract}
\maketitle

\section{Introduction}

The Fenna-Matthews-Olson (FMO) protein is one of the simplest and most 
well-studied light harvesting systems.
The protein complex exists as a trimer of three identical subunits
whose function is to link the chlorosome antenna complex,
where light-harvesting takes place, 
with the reaction center, where charge separation occurs. 
FMO is also one of the earliest light harvesting systems for which
a high resolution crystal structure has been obtained.\cite{fenna75}
This early crystal structure indicated that each of the three FMO subunits 
contains seven Bacteriochlorophyll (Bchl) chromophores
which serve as the primary energy transfer pathway between the 
chlorosome and the reaction 
center.\cite{brixner05,engel07,panitchayangkoon10,collini10}
Recently, however, a more careful crystallographic analysis of 
of FMO has been performed which demonstrates that the individual subunits 
contain eight Bchls, not seven.\cite{ben-shem04, tronrud09}
The eighth Bchl resides on the surface of the protein complex 
and it has been suggested that this additional chromophore is often lost
during sample preparation.

From an energy transfer perspective, the presence of an additional
chromophore may challenge current understanding of how exciton transfer
occurs in FMO. 
For example, in many previous studies on the seven Bchl complex, 
it is thought that two nearly independent energy transfer pathways 
exist.\cite{ishizaki09, huo10, tao10, wu10, wu11}
Sites one and six are approximately equidistant from the antenna complex
and both are assumed to be possible locations for accepting the 
excitation from the chlorosome.
From there, the energy is subsequently funneled either from site one to two
(pathway 1) or from site six to sites seven, five, and four (pathway 2).
The terminal point through either route is site three where the exciton
is then transferred to the reaction center 
(see~Fig.~\ref{fig:energy_diagram}(a)).
The couplings within each of the pathways are much larger than
the couplings between the two which implies that the two routes 
are nearly independent.\cite{ishizaki09a, huo10, wu11}
In the second paper of this series,
the two pathways and the respective probability of traversing 
each have been quantified using a flux analysis.\cite{wu11}

However, this two pathway picture is not entirely consistent 
with the recent experimental data.
The new crystal structure indicates that the eighth chromophore 
resides roughly midway between the baseplate and the 
Bchl at site one.\cite{ben-shem04, tronrud09, schmidt_am_busch11}
Additionally, Renger and coworkers argued in Ref.~\citenum{schmidt_am_busch11} 
that the eighth Bchl provides the most efficient path for exciton injection 
into FMO as a result of its position and orientation with respect
to the chlorosome.
If this is correct and site eight serves as the primary acceptor of 
excitation energy from the chlorosome,
then a preferential energy transfer route emerges through pathway 1. 
Due to the weak inter-pathway couplings, the secondary channel involving the
remaining four chromophores in pathway 2 is largely bypassed in this scheme.
This observation may have a significant impact on the efficiency and 
robustness of the energy transfer process.
The main objective of this work is to address this issue by exploring
how the dynamics and the energy transfer efficiency in FMO are affected by the 
presence of the eighth site and a realistic environment.

The first major conclusion of the present study is related to 
the population dynamics in FMO.
In many of the the previous studies of the seven site model of FMO,
the population relaxation dynamics are modeled with site one or 
site six initially populated.\cite{ishizaki09, huo10, tao10, wu10}
Under either of these initial conditions, pronounced oscillations in the
short-time dynamics are observed.
However when site eight is initially excited,\cite{schmidt_am_busch11} 
then the oscillations in the populations are completely suppressed.
This lack of oscillations has been independently observed in the dynamics 
recently reported in Refs.~\citenum{schmidt_am_busch11,renaud11,ritschel11}.
Here, we provide a simple explanation for this behavior.
The eighth Bchl maintains a large energy gap 
with the other seven sites in FMO in order to facilitate efficient 
directed energy transport.
However, it is also rather weakly coupled to the remaining Bchls.
This leads to a slow incoherent decay of the initial population at site eight,
and hence a broad distribution of initial conditions at the dimer.
The consequence of this result is that the population oscillations 
generally observed between sites one and two are completely suppressed
which illustrates the importance of the initial conditions on the 
dynamics of the dimer.

The second key result of this work demonstrates 
that if the eighth Bchl is the primary acceptor of excitation
energy from the chlorosome as recently proposed then a primary energy 
transfer pathway in FMO does indeed emerge.\cite{schmidt_am_busch11}
Note that this situation is substantially different from the previous 
interpretations of the energy flow in the seven site models where two
independent pathways are generally assumed to exist.
The extent of this effect and its impact on the energy transfer efficiency
is quantified by introducing reduced models of FMO that consist of only 
a subset of the sites in the full system.
It is demonstrated that sites eight, one, two and three which constitute
pathway 1 provide the largest contribution to the dynamics of the full system.
The remaining four sites of pathway 2 are seen to play a relatively small role.
Despite the fact that only a single pathway dominates the 
energy transfer process, we also show that the presence of the extra Bchl 
does not significantly impact the efficiency or robustness of FMO. 
The eight site model leads to only a slight increase in the transfer time 
as compared with the seven site system, and thus maintains
the same high efficiency as observed in previous studies of FMO.

Based upon energetic arguments, it has been suggested 
that the presence of the eighth Bchl 
leads to optimal energy transfer in FMO.\cite{schmidt_am_busch11}
That is, its location near the chlorosome allows for a large coupling 
to the antenna complex as well as substantial overlap of the absorption 
spectrum of the eighth Bchl with the fluorescence spectrum of the chlorosome.
These factors result in efficient transfer of the excitation energy into FMO
while simultaneously allowing the eighth chromophore to maintain 
a large energy gap with the remaining Bchls and hence a favorable
energy transport landscape.
These features implicitly suggest that there should be an optimal value 
of the site energy of the eighth Bchl.
Here it is demonstrated that this observation is correct.
However, in this case, the behavior is independent of the presence 
of the chlorosome, and can be understood by considering a further reduction 
pathway 1 to only three sites.
The result of this procedure is a downhill configuration of three 
equally spaced sites (see~Fig.~\ref{fig:energy_diagram}(c)) 
which is known to allow for highly efficient energy transfer.\cite{cao09}

Recently, several studies have shown that the environment does not have an
entirely destructive role in the energy transport properties of 
excitonic systems.\cite{mohseni08, wu10, wu11, plenio08, chin10}
Instead, the environment can serve to enhance both the efficiency and 
robustness of the energy transfer process.
Optimal values have been shown to exist for the temperature as well 
as other bath parameters which maximize the energy transfer efficiency in
several light harvesting systems.
Moreover, the experimentally fitted model parameters for FMO are near optimal 
in many cases. 
An extensive search for the optimal environmental parameters has been
recently presented in Refs.~\citenum{wu10} and~\citenum{mohseni11}.
In addition to the above findings, we also explore the effect of the eighth
Bchl on the environmentally assisted energy transport properties in FMO.
It is found that the optimal values of the bath parameters 
are similar to those found in the seven site model, but are 
closer to the experimental values in general.
As has been observed before, the energy transfer efficiency is 
relatively stable over a broad range parameters illustrating the 
robustness of the network.\cite{wu10}

In the next section, we present the average trapping time formalism 
which is used in the remainder of the discussion as a measure of the energy
transfer efficiency.\cite{cao97,wu10}
This is followed by a brief outline of the generalized Bloch-Redfield (GBR) 
approach and the model Hamiltonian for the eight site FMO complex 
used in the numerical calculations.
The results for the population dynamics and the development of the 
reduced models for FMO are then presented in Sec.~\ref{sec:results}.
This is followed by calculations of the trapping time as a function 
of the site energies and bath parameters.
There it is demonstrated that optimal values exist for many of these
parameters, and additionally that the experimentally fitted values for FMO
are near-optimal.

\section{Methods}\label{sec:methods}

\subsection{Average Trapping time}

The formalism for calculating the averaging trapping
time in light harvesting systems has been presented 
in detail previously in Refs.~\citenum{cao09} and~\citenum{wu10}.
Here we provide only the salient results.
The total system is characterized by a discrete $N$-site system 
Hamiltonian, $H_s$ and its interaction with the environment, $H_{sb}$.
Each site of the system is coupled to an independent bath of 
harmonic oscillators with the respective Hamiltonians,
$H_b = \frac12 \sum_j\left( p_j^2 + \omega_j^2 x_j^2\right)$.
The total Hamiltonian is then given by,
\begin{equation}
   H = \sum_n^N \epsilon_n \left| n \right \rangle \left \langle n \right |
     + \sum_{n\neq m}^N V_{nm} \left| n \right \rangle \left \langle m \right |
     + \sum_n^N \left| n \right \rangle \left \langle n \right |
       \left[H_b^{(n)} + \sum_j c_{j}^{(n)} x_{j}^{(n)} \right]
          \;,
\end{equation}
where $c_j^{(n)}$ denotes the coupling coefficient of site $n$
to the $j$-th mode of its associated bath.
The values of the site energies, $\epsilon_n$, and couplings 
constants, $V_{nm}$, are specified below in~Sec.~\ref{sec:FMO_model}.

The time evolution of the reduced density matrix of this system can be 
conveniently described in the Liouville representation as,
\begin{align} \label{eq:L_tot}
   \frac{\partial \rho(t)}{\partial t} & =  - L_{\rm tot} \rho(t) \nonumber \\
   & =  
     - \left( L_{\rm s} + L_{\rm trap} + L_{\rm decay} + L_{\rm sb} \right) 
     \rho(t) \;, 
\end{align}
where $L_{\rm s} \rho = i/\hbar \left[ H_s, \rho \right]$
describes the coherent evolution under the bare system Hamiltonian $H_s$.
In light harvesting systems, the energy flows irreversibly to the 
reaction center which is modeled here through the trapping operator
$[L_{\rm trap}]_{nm,nm} = \left( k_{t_m} + k_{t_n} \right)/2$,
where $k_{t_n}$ denotes the trapping rate at site $n$.
The energy transfer in FMO exhibits almost unit quantum yield.
As a result, the decay rate of the excitation at any site 
to the ground state, $k_d$, is expected to be much smaller than the 
trapping rate, $k_t \gg k_d$. 
This allows for the simplification $L_{\rm decay} = 0$.

\subsection{Generalized Bloch-Redfield Equation}\label{sec:GBR}

It remains to account for the Liouville operator describing the system-bath
coupling $L_{\rm sb}$ in~Eq.~\ref{eq:L_tot}.
For a harmonic bath linearly coupled to the system, the time correlation
function of the bath coupling operators is given by the 
standard relation\cite{Weiss99, mukamel99}
\begin{equation}
   C(t) = \frac1\pi \int_0^{\infty} d\omega \; 
   J(\omega)\left(\coth\left(\frac{\hbar \beta \omega}{2}\right)\cos(\omega t) 
             - i \sin(\omega t) \right) 
    \;,
\end{equation}
where $\beta=1/k_{\rm B}T$ and 
$J(\omega) = \frac{\pi}{2} \sum_j\frac{c_j^2}{\omega_j}\delta(\omega-\omega_j)$
is the spectral density of the bath.
For simplicity, we assume the spectral density is the same for each of 
the independent baths and given by the Drude form 
\begin{equation}\label{eq:drude}
   J(\omega) = 2\lambda \omega_c
             \frac{\omega }{\omega^2 + \omega_c^2}
   \;, 
\end{equation}
where $\lambda$ is the bath reorganization energy and $\omega_c$ is 
the Debye cutoff frequency.
For this special choice, the correlation function may be expanded 
in terms of the Matsubara frequencies, $\nu_j = \frac{2\pi j}{\hbar \beta}$,
as\cite{mukamel99, Weiss99,  wu10}
\begin{align} \label{eq:bath_corr}
   C(t) &= \left(
           \frac{2\lambda}{\hbar \beta} + \frac{4 \lambda\omega_c}{\hbar \beta}
           \sum_{j=1}^\infty \frac{\omega_c}{\omega_c^2-\nu_j^2}
           - i \lambda \omega_c 
           \right)
           e^{-\omega_c t}
           - \frac{4 \lambda \omega_c}{\hbar \beta}
           \sum_{j=1}^\infty \frac{\nu_j}{\omega_c^2 - \nu_j^2}e^{-\nu_j t}
            \notag \\
   & = \sum_{j=0}^\infty \alpha_j e^{-\nu_i t}
   \;, 
\end{align}
which defines the complex expansion coefficients $\alpha_j$
with the condition $\nu_0 = \omega_c$.

The dynamics in FMO have been computed using a variety of methods ranging 
in both accuracy and 
cost.\cite{ishizaki09a, huo10, tao10, renaud11, mohseni11, pachon11}
Here, we choose the approximate generalized Bloch-Redfield (GBR) method 
which follows from a second order cumulant expansion in the 
system-bath interaction.
It provides an accurate, but computationally friendly approach for the
propagation of the density matrix over much of the physically relevant 
parameter space.\cite{cao97, wu10}
Due to the decomposition of the bath autocorrelation function in 
Eq.~\ref{eq:bath_corr}, the system-bath interaction may be accounted for
through the introduction of auxiliary fields.
The GBR equation of motion for the reduced density matrix is then given by
\begin{equation}
   \frac{\partial \rho(t)}{\partial t} =
                  -\left(L_{\rm sys} + L_{\rm trap}\right)\rho(t)
   - i \sum_n^N \sum_{j=0}^\infty \left[A_n, g_{n,j}(t)\right]
   \;.
\end{equation}
The coupling of each Bchl to an independent bath leads to the 
additional sum over the $N$ sites where the system operator
$A_n = \left| n \right \rangle \left \langle n \right |$ and
$g_{n,j}$ denotes the $j$th-auxiliary field coupled to site $n$.
The auxiliary variables are subject to the initial 
conditions $g_{n,j}(0) = 0$ and obey the equations of motion,
\begin{equation}
   \frac{\partial  g_{n,j}(t)}{\partial t} =
     -\left(L_{\rm sys}+L_{\rm trap} + \nu_j\right) g_{n,j}(t)
   - i {\rm Re}(\alpha_j) \left[A_n,\rho(t)\right] 
   +  {\rm Im}(\alpha_j) \left[A_n,\rho(t)\right]_{+}
    \;,
\end{equation}
where the plus subscript denotes anti-commutation.

\subsection{Trapping Time}

The mean residence time at site $n$ is by definition 
\begin{equation}
   \langle \tau_n\rangle = \int_0^\infty dt \rho_{nn}(t)
   \;,
\end{equation}
where $\rho_{nn}$ denotes the population at site $n$.
The average trapping time is then given simply as the sum of the residence
times at each of the $N$ sites,
$\langle t \rangle = \sum_{n=1}^N \langle \tau_n \rangle$.
Invoking the steady state solution of~Eq.~\ref{eq:L_tot}, 
$L_{\rm tot} \langle t \rangle =  \rho(0)$, then the 
average trapping time is given by the compact expression,
\begin{equation} \label{eq:trap_time}
   \langle t \rangle = 
      {\rm Tr} \left(L_{\rm tot}^{-1}\rho(0)\right)
   \;, 
\end{equation}
where the trace is taken over the site populations of the reduced
density matrix.

\subsection{Eight Site FMO model}\label{sec:FMO_model}

The Hamiltonian for FMO is constructed from the crystal structure 
recently deposited in the protein data bank (pdb code: 3eoj).\cite{tronrud09}
The site energies are taken from those computed in 
Ref.~\citenum{schmidt_am_busch11} and the coupling element
between sites $n$ and $m$ is calculated from the dipole-dipole approximation,
\begin{equation}
   \label{eq:dipole}
   V_{nm} = C \left(\frac{\mathbf d_n \cdot \mathbf d_m}
                   {\left| \mathbf r_{nm}\right|^3}
   -3 \frac{\left(\mathbf d_n \cdot \mathbf r_{nm}\right) 
            \left(\mathbf d_m \cdot \mathbf r_{nm}\right)  }
           {\left| \mathbf r_{nm}\right |^5} \right)
           \;. 
\end{equation}
Additional details and the explicit system Hamiltonian are given in 
the Appendix.
Aside from the eighth site, the most significant difference between 
the present model Hamiltonian 
and the model previously derived by Cho et al.\cite{cho05} is in the 
energy difference between sites one and two.
In the current  case, the energy transfer through pathway 1 is entirely
energetically favorable whereas a barrier is present between site one and two
in the model of Ref.~\citenum{cho05}.

Unless otherwise stated, the bath is characterized by the experimentally
fitted values for the reorganization energy of $35$ cm$^{-1}$ and Debye 
frequency $\omega_c^{-1}=50$ fs ($105$ cm$^{-1}$).\cite{cho05, ishizaki09}
Additionally the temperature is $300$ K and the trap is located 
at site three with a trapping rate of $k_t = 1$ ps$^{-1}$.

\section{Numerical Results}\label{sec:results} 

\subsection{Population Dynamics and suppression of the oscillations }

\begin{figure}
   \includegraphics*[width=\textwidth]{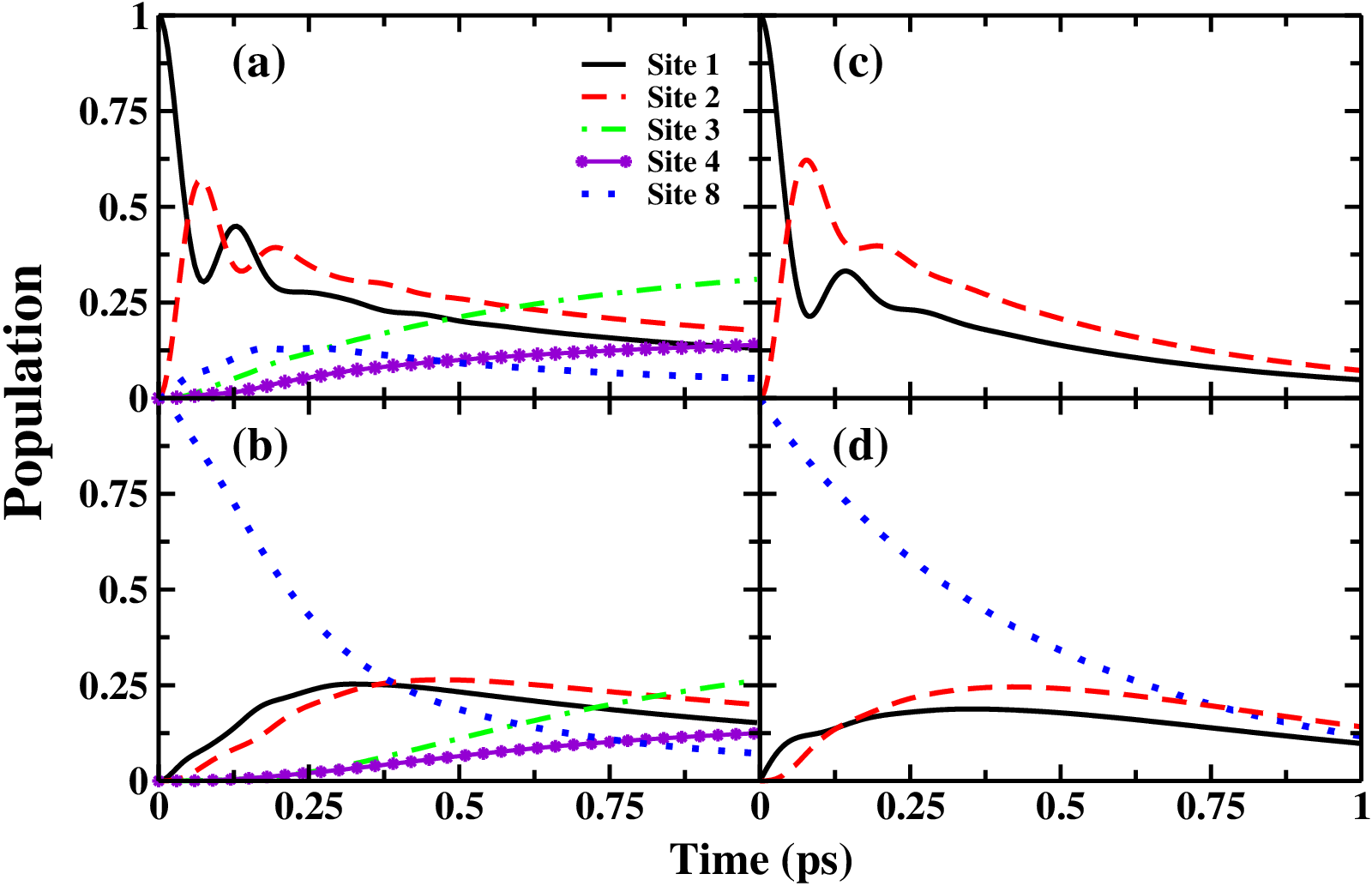}
   \caption{
   Site populations of in the eight site model of FMO with Bchl one 
   (a) or eight (b) initially excited calculated with the GBR.
   The populations of the remaining sites five, six and seven are never
   larger than $10\%$ and not shown.
   The site populations of the dimer calculated using the NIBA of~Eq.~\ref{eq:niba}
   calculated with site one initially excited (c) and 
   from~Eq.~\ref{eq:convolution} (d).
   In all cases, the temperature is $300$ K with a reorganization energy of 
   $35$ cm$^{-1}$ and cutoff frequency of $\omega_c^{-1} = 50$ fs.
   }\label{fig:P_site}
\end{figure}

The time evolution of the populations in the eight site model of FMO 
calculated from~Eq.~\ref{eq:L_tot} using the GBR
is shown in Fig.~\ref{fig:P_site}(a) and (b)
for the initial population located at site one and site eight, respectively.
The bath parameters are taken at their fitted values specified above
and the trap at site three is not included.
The most striking difference seen between Fig.~\ref{fig:P_site}(a) and (b)
is the absence of the oscillations in the populations of the dimer 
when site eight is initially excited.
Other recent studies of the eight site model of FMO have also observed 
a similar lack of oscillations.\cite{schmidt_am_busch11,renaud11,ritschel11}
The origin of this effect may be traced to the initial conditions at
the dimer.
The energy difference between site eight and the remaining sites
is much larger than any of its respective couplings.
This leads to the rather slow incoherent exponential relaxation of the 
population of site eight seen in~Fig.~\ref{fig:P_site}(b).
The resulting initial conditions at sites one and two are then given
by a corresponding incoherent distribution.
It is this dephasing that suppresses the oscillations generally
observed in the dynamics of the dimer.

By applying the non-interacting blip approximation (NIBA) to the spin-boson 
model, Pachon and Brumer established that a necessary condition
for the presence of the oscillations in the dimer is an effective 
low temperature.\cite{pachon11}
The results of~Fig.~\ref{fig:P_site} demonstrate that the initial conditions
impose an additional constraint on the observation of population oscillations.
Note that there are a variety of other initial preparations
--such as starting from an eigenstate of the total system or exciting
the system with incoherent light\cite{brumer11}--
which will also suppress the oscillations in the dimer.

In order to analyze the influence of the initial conditions in more detail,
the dynamics of the dimer calculated using the NIBA 
are presented in~Fig.~\ref{fig:P_site}(c) and (d).
The population dynamics described in Ref.~\citenum{pachon11} may be 
formulated as an equivalent generalized master 
equation\cite{silbey84, aslangul86, cao00}
\begin{equation}
   \frac{\partial  P_n(t)}{\partial t} = \int_0^t dt^\prime\; 
                 \sum_{m=1}^N K_{nm}(t-t^\prime) P_m(t^\prime)
   \;, \label{eq:niba}
\end{equation}
where $P_n(t)$ denotes the population of site $n$ at time $t$.
The elements of the time-dependent transition matrix are 
constructed in the standard fashion\cite{zwan01}
\begin{equation}
   K_{nm}(t) = \left( 1 - \delta_{nm}\right) W_{mn}(t) 
             - \delta_{nm} \sum_k W_{nk}(t)
             \;
\end{equation}
where $\delta_{nm}$ is the Kronecker delta function and 
the individual rate kernels are given by the NIBA
\begin{equation}
   W_{nm}(t) = 2\: V_{nm}^2\:  
             e^{i\left(\epsilon_n - \epsilon_m - 2\lambda\right) t - 2 C(t) }
   \;.
\end{equation}
As defined previously, the coupling between site $n$ and $m$ is denoted 
by $V_{nm}$, $\epsilon_n$ is the energy of site $n$, 
$\lambda$ is the reorganization energy, and 
$C(t)$ is the bath correlation function given in~Eq.~\ref{eq:bath_corr}.
The results shown in~Fig.~\ref{fig:P_site}(c) are calculated from~Eq.~\ref{eq:niba}
with the initial population located at site one and are seen to capture the 
key features of the full GBR dynamics shown in~Fig.~\ref{fig:P_site}(a).
The decay is accounted for by setting the transfer elements 
$K_{n1}$ and $K_{n2}$ to zero for all sites $n>2$
which allows for population transfer from the dimer to the remaining
Bchls, but prevents any back-transfer.
Effectively this results in the addition of traps at sites one and two
and thus leads to the population decay of the dimer.
Without this decay, the two-site dynamics reproduce those of 
Ref.~\citenum{pachon11}.

As has been alluded to previously, the lack of oscillations in the dimer
when the initial population is located at site eight may be explained by 
creating a distribution of initial conditions at site one.
All of the population is initially located at site one 
in Fig.~\ref{fig:P_site}(a), 
whereas in~Fig.~\ref{fig:P_site}(b), the corresponding initial conditions 
are given by the population that steadily flows from site eight.
This slow incoherent decay of the population at site eight 
may be adequately fit to the single 
exponential $P_8(t) = \exp(-\gamma t)$ where $\gamma = 3$ ps$^{-1}$.
Assuming that site eight decays only into site one, then
the population of the latter is given by $1-P_8(t)$,
and the corresponding transition rate is $W_{81}(t) =  \gamma \exp(-\gamma t)$.
The influence of site eight on the oscillatory behavior of the dimer 
can then be captured by convoluting the dynamics given in~Fig.~\ref{fig:P_site}(c)
(denoted by $P_n(t)$) with this initial condition,\cite{kubo85} 
\begin{equation}
   \bar P_n(t) = \int_0^t dt^\prime\; P_n(t-t^\prime) W_{81}(t^\prime) \;.
   \label{eq:convolution}
\end{equation}
The result of this procedure is shown in~Fig.~\ref{fig:P_site}(d).
As is evident, the oscillatory behavior has completely disappeared.
For large $\gamma$, the transition rate becomes a delta function 
and the dynamics of~Fig.~\ref{fig:P_site}(c) are recovered.
However, oscillations in the populations of both states can
be observed only if the decay rate out of site eight is increased fivefold.
The presence or absence of the trap simply effects the long time decay 
of the dimer populations and is irrelevant for the short time 
oscillatory behavior.
These results demonstrate the importance of the initial preparation
of the populations on the oscillations in the dynamics.

It should be noted that while the NIBA calculations lead to qualitatively
similar population dynamics as those given by the GBR, neither of the two 
approaches are exact.
In FMO and other light harvesting systems, many of the model parameters
are of the same order of magnitude.
For instance, the couplings, $V_{nm}$, and energy 
differences, $\epsilon_n - \epsilon_m$,
as well as the reorganization energy, $\lambda$, 
and thermal energy, $\beta^{-1}$, are all of comparable magnitude.
As a result, methods based upon second-order perturbation theory are, 
in general, insufficient to quantitatively describe the dynamics.
A systematic procedure for computing higher-order contributions to the
NIBA rates has been derived and recently implemented.\cite{cao00,wu11}
This leads to non-negligible corrections to the dynamics in the spin-boson
model, FMO and LH2.\cite{wu12}
Thus while the results in~Fig.~\ref{fig:P_site} 
and those of Ref.~\citenum{pachon11} capture the 
qualitative features that are necessary to analyze the energy transfer behavior,
there will be quantitative corrections from higher order terms.

\subsection{Pathway analysis and the ladder configuration}

Regardless of the presence or absence of oscillations, it is readily
seen from~Fig.~\ref{fig:P_site}(a)~and~(b) that the population primarily 
flows through pathway 1.
Among the sites in pathway 2, only site four ever accumulates more than 
ten percent of the population.
Particularly for times less than $500$ fs, the sites from pathway 2 
are scarcely populated.
Similar behavior of the population dynamics has been seen 
in many other simulations of the seven site model for 
FMO.\cite{schmidt_am_busch11, renaud11, ritschel11, ishizaki09a, huo10, tao10}
These observations lead to the first reduced model for FMO which 
consists of only 
the four sites in pathway one shown in~Fig.~\ref{fig:energy_diagram}(b).
As demonstrated below, this model is able to accurately capture the 
key features of the energy transfer process.
One may proceed further by noting that sites one and two are coupled 
more strongly than the energy difference between the two.
Additionally, sites eight and three are widely separated from either site in the
dimer.
The couplings between the distant sites and dimer are also rather weak
(see values in the model Hamiltonian in 
Eq.~\ref{eq:Ham} and~Fig.~\ref{fig:energy_diagram}(b)).
As a result, there can be rapid coherent energy transfer between sites 
one and two, whereas the transfer to sites eight or three will 
be comparatively slow.
Therefore, when the initial population is located at site eight 
we may simply assume that these two sites of the dimer behave as one 
effective site with an energy that is the average of the two ($270$ cm$^{-1}$).
A similar ``mean state'' idea for developing this type of reduced model
has been suggested in Ref.~\citenum{huo11} which explores the behavior of a
dimer embedded in the PC645 photosynthetic network.
Furthermore, the coupling between site eight and the terminal site, three,
is negligibly small.
This leads to a three site model for FMO where the couplings are 
determined by the nearest-neighbor values as
shown in~Fig.~\ref{fig:energy_diagram}(c).

\subsection{Site Energy of Bchl 8}

\begin{figure}
   \includegraphics*[width=0.95 \textwidth]{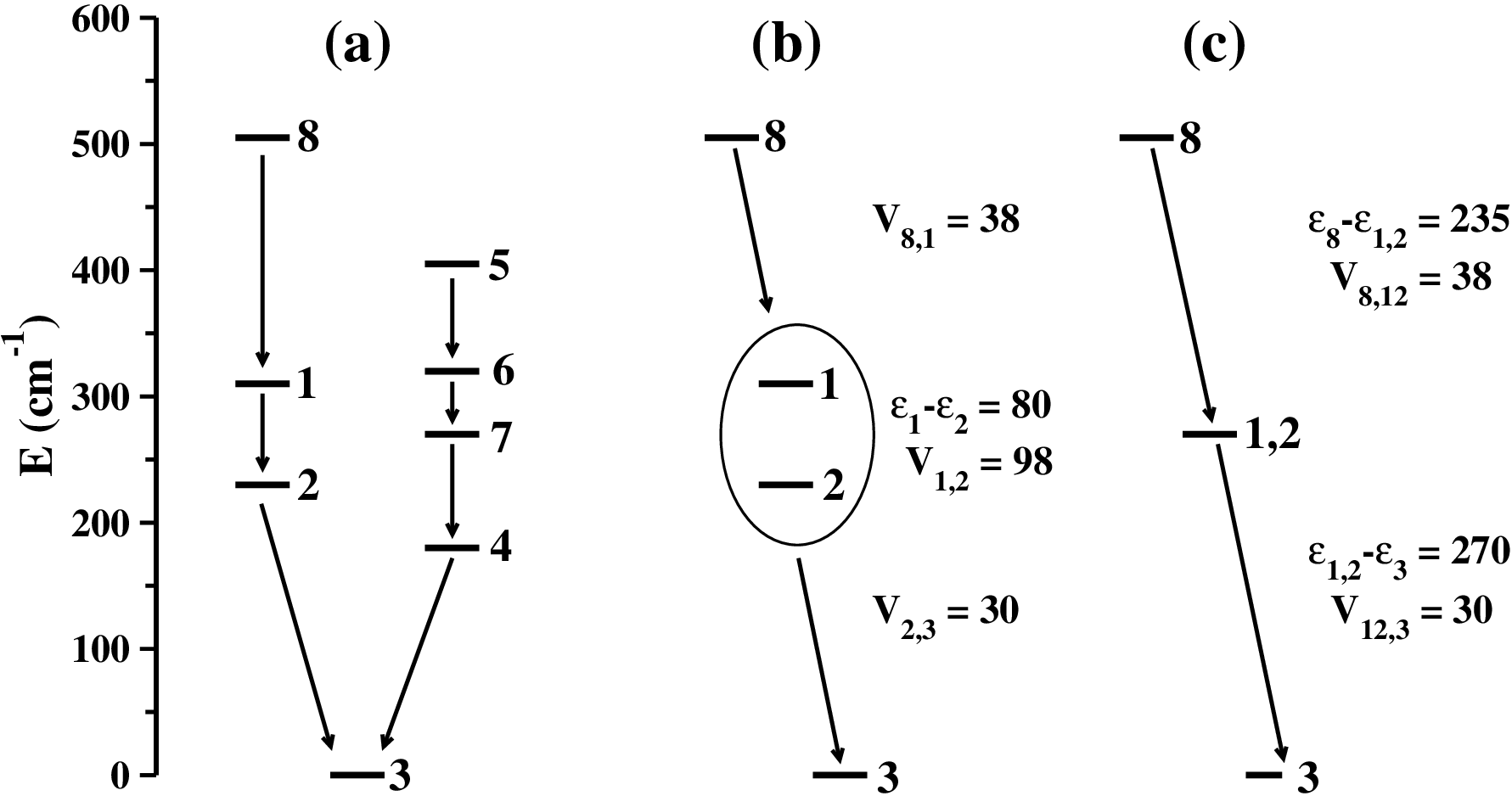}
   \caption{
   Energy diagrams for the eight site model (a), the four site model (b), 
   and three site model (c) used in the calculations 
   of~Fig.~\ref{fig:site_energy}.
   }\label{fig:energy_diagram}
\end{figure}

\begin{figure}
   \includegraphics*[width=0.75 \textwidth]{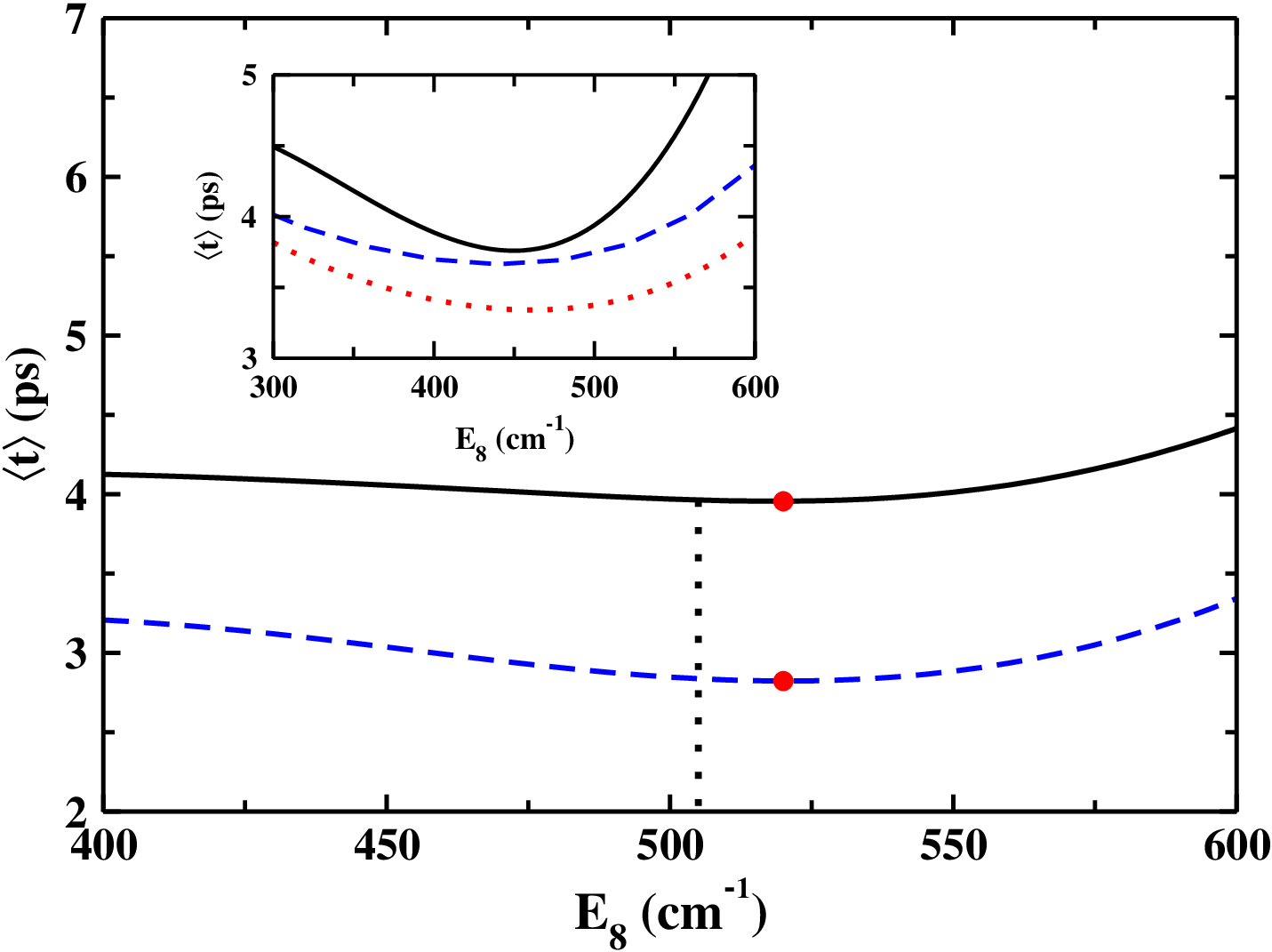}
   \caption{
   The trapping time as a function of the site energy of site eight.
   The solid (black) line and dashed (blue) line in the main figure 
   correspond to the results calculated from the full 
   eight site Hamiltonian of~Eq.~\ref{eq:Ham}
   and with only the four sites of pathway 1, respectively.
   The red dots correspond to the optimal site energies and the 
   vertical dashed line indicates the fitted value of the site energy 
   of Bchl 8 of $505$ cm$^{-1}$ determined in Ref.~\citenum{schmidt_am_busch11}.
   The inset contains results for the three site model
   calculated with the GBR (solid black line), F{\"o}rster theory 
   (dotted red line), and hierarchical equation of motion (dashed blue line).
   The remaining parameters are the same as in~Fig.~\ref{fig:P_site}.
   }\label{fig:site_energy}
\end{figure}

Fig.~\ref{fig:site_energy} displays the average trapping time calculated
as a function of the site energy of Bchl eight. 
It contains two additional key findings of this work.
The first is that the trapping time behavior, and hence the efficiency,
seen in the eight site model of FMO is largely governed by the sites 
in pathway 1.
The second feature is that an optimal value exists for the site energy 
of Bchl 8, and moreover, the optimum is near the fitted value determined in 
Ref.~\citenum{schmidt_am_busch11}.
The source of the latter is the highly efficient ladder 
configuration shown in~Fig.~\ref{fig:energy_diagram}(c).

The main portion of~Fig.~\ref{fig:site_energy} displays the average trapping 
time calculated with the full Hamiltonian given in~Eq.~\ref{eq:Ham} 
along with the corresponding results for the four site model of FMO 
(see~Fig.~\ref{fig:energy_diagram}(a) and (b)).
The bath parameters are taken at their experimentally fitted values 
with the temperature of $300$ K.
The inset of~Fig.~\ref{fig:site_energy} contains the results from 
the three sites model shown in~Fig.~\ref{fig:energy_diagram}(c).
For this case, the exact trapping time calculated from the hierarchical 
equation of motion method is also presented,
as well as the results of a F{\"o}rster theory calculation.
As can be seen, both the GBR and the F{\"o}rster calculations predict 
optimal values of the site energy that semi-quantitatively 
capture the behavior of the exact hierarchical results.

These results demonstrate that the energy transport is dominated by 
a subset of the sites in FMO and furthermore that the mechanism 
is correctly described by F{\"o}rster theory.
The four site model correctly describes the qualitative features seen in the
full system and it accounts for the majority of the trapping time.
Of the remaining sites in~Fig.~\ref{fig:P_site}, site four was seen to 
have the largest impact on the population relaxation.
Calculations that consist of pathway 1 plus site four are seen to 
capture almost all of the behavior seen in the full eight site system.

In Ref.~\citenum{schmidt_am_busch11}, it was noted that the site energy 
of Bchl eight maximizes the overlap with the chlorosome emission spectrum 
while simultaneously maintaining a large energy gap with the remaining 
seven core Bchls.
This indicates that there should be an optimal value of the site energy
of Bchl eight.
In addition to the observation that the trapping time behavior may
be captured by a simplified model of FMO, there is another
interesting feature seen in~Fig.~\ref{fig:site_energy}.
The trapping time displays a minimum as a function of the energy of 
the eighth site for all of the constructed models.
Moreover, for the eight site model, the optimal value of the site energy
is rather close to the fitted value of 
$505$ cm$^{-1}$.\cite{schmidt_am_busch11} 
Increasing the energy difference between site eight and site one decreases the
back-transfer rate, but also decreases the spectral overlap between the two.
The position of the optimal value is no coincidence.
The three-site model in~Fig.~\ref{fig:energy_diagram}(c) readily 
demonstrates that this particular choice for the site energy leads 
to a downhill configuration of three sites that are approximately 
equally-spaced and equally-coupled 
which is known to be very efficient.\cite{cao09}
The qualitative behavior of the trapping time in the full, complicated 
eight site system becomes obvious with the aid of the reduced models.
Note also that the average trapping time varies little over a 
large range of values of the energy of site eight indicating 
the robustness of the energy transfer process.
Below it is demonstrated that many of the other fitted parameters are near
optimal as well.

\subsection{Optimal Bath Parameters}

\begin{figure}
   \includegraphics*[width=\textwidth]{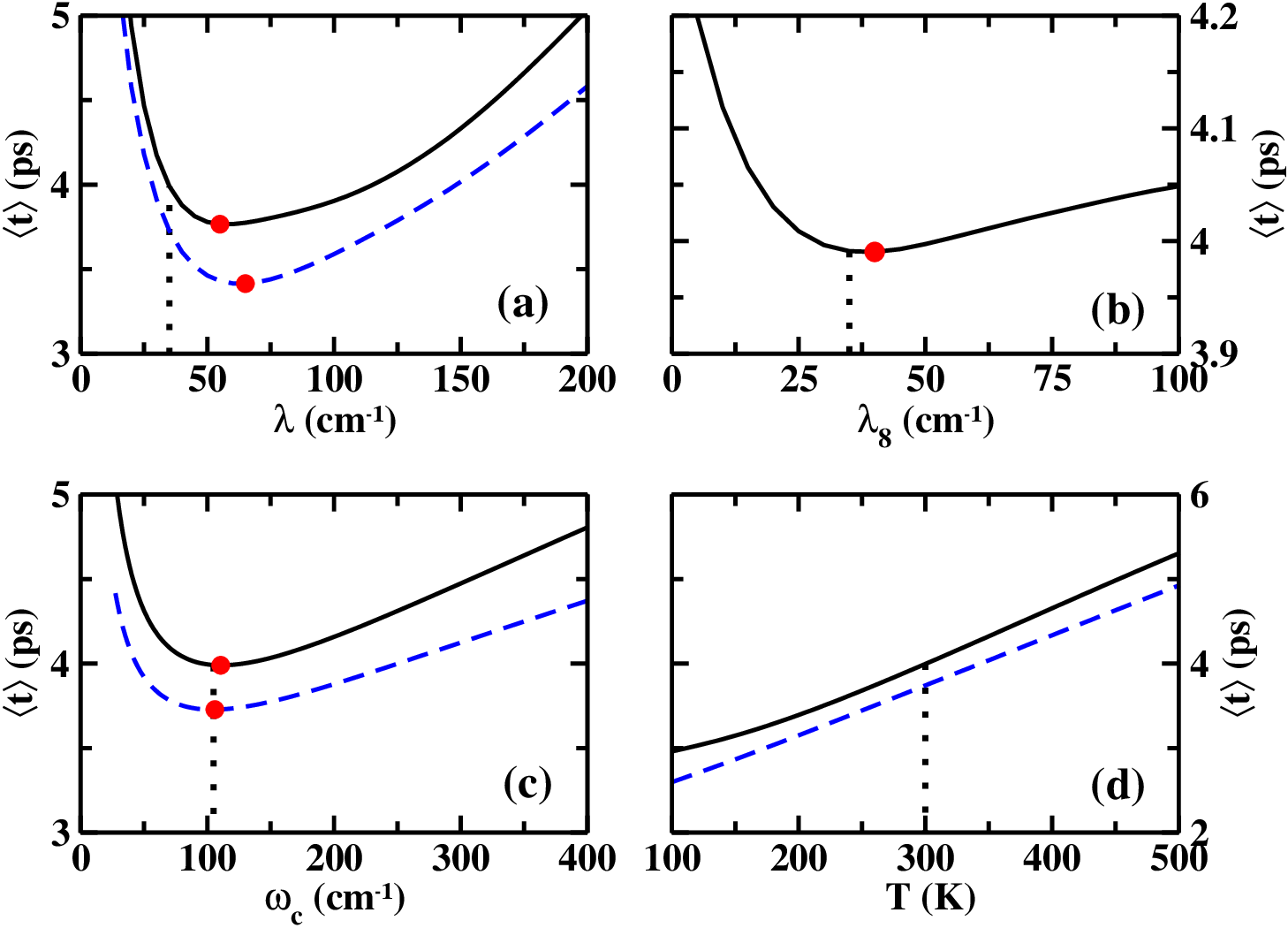}
   \caption{
   The trapping time as a function of the reorganization energy of 
   all Bchls (a), and as a function of the reorganization of site eight 
   only (b) while the remaining seven sites are fixed at the experimentally
   fitted value of $\lambda=35$ cm$^{-1}$. 
   The trapping time as a function of the bath cutoff frequency 
   and temperature are shown in figures (c) and (d) respectively.
   In all cases the trap rate at site three is $1$ ps.
   The solid (black) and dashed (blue) lines correspond to initial 
   excitation at site eight or site one, respectively.
   The red dots indicate the optimal trapping times and the 
   dotted vertical lines correspond to the respective experimentally 
   fitted values.
   The remaining parameters are the same as in~Fig.~\ref{fig:P_site}.
   }\label{fig:bath}
\end{figure}

The average trapping time calculated as a function of the reorganization
energy is shown in~Fig.~\ref{fig:bath}(a) and (b).
Fig.~\ref{fig:bath}(a) varies the reorganization of all eight sites
simultaneously 
(in this model all chromophores are assumed to have identical environments) 
whereas (b) varies only that of site eight while keeping
all of the others at the fitted value of $35$ cm$^{-1}$.
In order to demonstrate that the presence of the additional chromophore
does not lead to a dramatic increase in the trapping time, 
two different initial conditions are taken with either site one or site eight 
initially excited.
As expected, a slightly larger trapping time is observed with the 
initial population at site eight due to the larger distance to the trap.
However, the difference between the two scenarios is not substantial.
Additionally, the qualitative behavior of the two initial conditions
is quite similar and both lead to an optimal value of the reorganization
energy that is close to the experimentally fitted value of $35$~cm$^{-1}$.
It has been proposed from recent numerical simulations that the 
reorganization energy in FMO should be approximately twice as large as 
the experimentally fitted value used here.\cite{olbrich11}
The optimal values of $\lambda$ 
in~Fig.~\ref{fig:bath}(a) ($55$ cm$^{-1}$) and (b) ($40$ cm$^{-1}$) 
are consistent with a somewhat larger value of the reorganization energy.
The mean trapping time is more sensitive to variations of 
the reorganization energy than was observed for the site energy 
in~Fig.~\ref{fig:site_energy}, but there is still a large range of 
$\lambda$ where the trapping time is near optimal.

Finally,~Fig.~\ref{fig:bath}(c) and (d) display the results for the 
average trapping time calculated as a function of the Debye frequency and 
as a function of the temperature.
As in~Fig.~\ref{fig:bath}(a) two initial conditions are shown with the
initial population at either site one or site eight.
Again, the average trapping time from site one is always faster than for 
site eight.
Nevertheless, the two initial conditions display qualitatively similar 
behavior for both the temperature and Debye frequency.
Additionally an optimal value of the Debye frequency is observed 
that is close to the experimentally fitted value of $105$ cm$^{-1}$.
These results for the average trapping time as a function of the 
bath parameters are similar to those observed previously for the seven site
model of FMO.\cite{wu10}
There is one notable difference however.
In all cases, the Hamiltonian for the eight site FMO model 
recently proposed by Renger et al\cite{schmidt_am_busch11} 
leads to optimal conditions that are closer to the experimentally 
fitted values than those calculated previously\cite{wu10} using 
the Hamiltonian of Ref.~\citenum{cho05}.

\section{Conclusions}

The FMO protein serves as one of the model light harvesting systems,
and the qualitative features of the energy transfer process have been
understood for some time.
However, recent experimental evidence has shown that many of 
the previously developed theoretical models are not entirely complete.
An additional chromophore is present in each subunit of FMO that 
resides between the chlorosome and site one.
In this work we have shown that the presence of the eighth site 
does not significantly alter the previous conclusions that have been
reached with regards to environmentally-assisted exciton transport.\cite{wu10}
Optimal values exist for many of the bath parameters and, moreover,
the optimal conditions are generally closer to their respective 
experimentally fitted values than in the seven site FMO models.
Additionally, the dependence of the average trapping time with respect
to variations of the bath parameters is rather weak illustrating the
overall robustness of the energy transfer process.

However, the presence of the eight Bchl may necessitate a 
reassessment of our understanding of the energy transport process in FMO.
Given that site eight is the primary entry point for the exciton into FMO,
then~Fig.~\ref{fig:P_site} clearly exhibits a complete suppression 
of the population oscillations that are generally observed in the seven site
models of FMO.
That is, the coherent population oscillations observed in previous studies
depend upon the special choice of initial conditions.
Here we have shown that the origin of the suppression is the slow decay 
of the initial population at site eight which leads to an incoherent 
distribution of initial conditions at the dimer.
In the physical setting there will be an additional source of 
dephasing due to the extra step from the chlorosome to FMO.
This will broaden the distribution of initial conditions even more
and further suppress the oscillations in the dimer.

An additional feature of the eight site model that is markedly different
from the previous seven site configuration is observed in the 
population dynamics shown in Fig.~\ref{fig:P_site} and the average trapping 
times displayed in~Fig.~\ref{fig:site_energy}.
These results demonstrate that the energy flow in the eight site model 
is dominated by a subset of the chromophores, 
whereas it has been previously assumed that two independent pathways 
involving all of the Bchls are available for the energy transfer process.
The qualitative features of the transport in the eight site model
are largely determined by the dynamics of pathway 1.
Sites four, five, six and seven provide a rather small contribution to the 
overall efficiency in this case.
The agreement between the results for the full eight site system 
and the reduced four- and three-site models shown in 
Fig.~\ref{fig:energy_diagram} provide further support to this claim.
Nevertheless, the eight site model and the seven site model display
similar energy transport efficiencies.
The origin of this behavior in the former is evident from
Fig.~\ref{fig:energy_diagram}(c) which shows that the 
eighth Bchl forms an optimal downhill ladder configuration with the dimer
and site three.
This result demonstrate the usefulness of the reduced models 
in providing an intuitive explanation of many of the key features 
present in the numerical results.

\section{acknowledgment}
This work was supported by grants from the National Science Foundation, 
DARPA, the Center for Excitonics at MIT, the MIT energy initiative (MITEI), and
the Singapore-MIT alliance for research and technology (SMART)

\section{Appendix}

Following the prescriptions used previously for constructing the dipole-dipole 
interactions,\cite{hu97, muh07, schmidt_am_busch11} 
the unit vectors, $\mathbf d_n$, in~Eq.~\ref{eq:dipole} point along the axis 
connecting the $N_b$ and $N_d$ atoms of the $n$-th Bchl 
and $\mathbf r_{nm}$ is the vector connecting the Mg atoms of Bchl $n$ and $m$.
Setting the constant $C=155000$ cm$^{-1} $\AA$^3$ leads to 
an effective dipole strength of 30 D$^2$.
With these specifications, the system Hamiltonian (in cm$^{-1}$) 
for the eight site model is 
\begin{equation}
   \label{eq:Ham}
   H_{\rm FMO} =
\begin{pmatrix}
  310.0 &   -97.9  &  5.5 &   -5.8 &   6.7 & -12.1 & -10.3 &  37.5 \\ 
  -97.9 &   230.0  & 30.1 &    7.3 &   2.0 &  11.5 &   4.8 &   7.9 \\
    5.5 &    30.1  &  0.0 &  -58.8 &  -1.5 &  -9.6 &   4.7 &   1.5 \\
   -5.8 &     7.3  &-58.8 &  180.0 & -64.9 & -17.4 & -64.4 &  -1.7 \\
    6.7 &     2.0  & -1.5 &  -64.9 & 405.0 &  89.0 &  -6.4 &   4.5 \\
  -12.1 &    11.5  & -9.6 &  -17.4 &  89.0 & 320.0 &  31.7 &  -9.7 \\
  -10.3 &     4.8  &  4.7 &  -64.4 &  -6.4 &  31.7 & 270.0 & -11.4 \\ 
   37.5 &     7.9  &  1.5 &   -1.7 &   4.5 &  -9.7 & -11.4 & 505.0 
\end{pmatrix}\;, 
\end{equation}
where the zero of energy is $12195$ cm$^{-1}$.
Note that there is an error in the sign of the coupling between sites
one and two in the table provided in Ref.~\citenum{schmidt_am_busch11}.
Aside from this, these values reproduce all of the couplings listed 
therein to within $3$ cm$^{-1}$.


\newpage

\end{document}